# A Chaotic Approach to Market Dynamics

Carmen Pellicer-Lostao[1], Ricardo López-Ruiz[2]

*Department of Computer Science and BIFI,*

*Universidad de Zaragoza, 50009 - Zaragoza, Spain.*

e-mail address: carmen.pellicer@unizar.es[1], rilopez@unizar.es[2]

**ABSTRACT:**

Economy is demanding new models, able to understand and predict the evolution of markets. To this respect, Econophysics is offering models of markets as complex systems, such as the gas-like model, able to predict money distributions observed in real economies. However, this model reveals some technical hitches to explain the power law (Pareto) distribution, observed in individuals with high incomes. Here, non linear dynamics is introduced in the gas-like model. The results obtained demonstrate that a "chaotic gas-like model" can reproduce the two money distributions observed in real economies (Exponential and Pareto). Moreover, it is able to control the transition between them. This may give some insight of the micro-level causes that originate unfair distributions of money in a global society. Ultimately, the chaotic model makes obvious the inherent instability of asymmetric scenarios, where sinks of wealth appear in the market and doom it to complete inequality.

# 1. INTRODUCTION:

After the economic crisis of 2008, much criticism has been thrown upon modern economic theories. They have failed to predict the financial crisis and foresee its depthless [1]. On the whole, this failure has been charged to the inherent complexity of the markets [2].

In this respect, Econophysics has been ultimately offering new tools and perspectives to deal with economic complexity [3-5]. Simple stochastic models have been developed, where many agents interact at micro level giving rise to global empirical regularities observed in real markets. These models are giving some guidance to uncover the underlying rules of real economy.

One of the most relevant examples of these models is the conjecture of a kinetic theory with (ideal) gas-like behaviour for trading markets [4-9]. This model offers a simple scheme to predict money distributions in a closed economic community of individuals. In it, each agent is identified as a gas molecule that interacts randomly with others, trading in elastic or money-conservative collisions. Randomness is also an essential ingredient, for agents interact in pairs chosen at random and exchange a random quantity of money. In the end, the model shows that the distribution of money in the community will follow the exponential (Boltzmann-Gibbs) law for a wide variety of trading rules [5].

This result may elucidate real economic data, for nowadays it is well established that income and wealth distributions show one phase of exponential profile that covers about 90-95% of individuals (low and medium incomes) [10-12].

Despite of this fact, the model shows some technical hitches to explain the other phase observed in real economies. This is the power law (Pareto) distribution profile integrated by the individuals with high incomes [12-15]. To obtain it, the model needs to introduce additional elements, such as saving propensity or diffusion theory [5, 9].

The work presented here, intends to contribute in this respect. To do that, it proposes to incorporate chaotic dynamics in the traditional gas-like models. This is done upon two facts that seem particularly relevant in this purpose.

The first one is that, determinism and unpredictability, the two essential components of chaotic systems, take part in the evolution of Economy and Financial Markets. On one hand, there is some evidence of markets being not purely random, for most economic transactions are driven by some specific interest (of profit) between the interacting parts. On the other hand, real economy shows periodically its unpredictable component with recurrent crisis. The prediction of the future situation of an economic system resembles somehow the weather prediction. Therefore, it can be sustained that market dynamics appear to be chaotic; in the short-time they evolve under deterministic forces though, in the long term, these kind of systems show an inherent instability.

The second fact is that the transition from the Boltzmann-Gibbs to the Pareto distribution may require the introduction of some kind of inhomogeneity that breaks the random indistinguishability among the individuals in the market [16]. Nonlinear maps can be an ideal candidate to obtain that, as they can produce quite simple routines of evolution for the market. In particular, they also represent a simple mechanism to introduce another important ingredient: some degree of correlation between agents. In this way chaotic dynamics is able of breaking the pairing symmetry of interactions and establishing a complex pattern of transactions.

These concepts or hypothesis have inspired the introduction of a model for trading markets with chaotic patterns of evolution. Amazingly, it will be seen that a chaotic market is able to reproduce the two characteristic phases observed in real wealth distributions. The aim of this work is to observe what may happen at a micro level in the chaotic market, which is responsible of producing these two global phases.

This paper is organized as follows: section 2 introduces the basic theory of the gas-like model and describes the simulation scenario used in the computer simulations. Section 3 shows the results obtained in these simulations. Final section gathers the main conclusions and remarks obtained from this work.

## 2. SIMULATION SCENARIO OF THE CHAOTIC MARKET

The model proposed here is a multi-agent gas-like scenario [5] where the selection of agents is chaotic, while the money exchanged at each interaction is a random quantity.

As in the gas-like model scenario a community of $N$ agents is given an initial equal quantity of money, $m_0$. The total amount of money, $M=N*m_0$, is conserved. The system evolves for a total time of $T=2*N^2$ transactions to reach the asymptotic equilibrium.

For each transaction, at a given instant $t$, a pair of agents $(i, j)$ with money $(m^t_i, m^t_j)$ is selected chaotically and a random amount of money $\Delta m$ is traded between them. The amount $\Delta m$ is obtained through equations (1) where $\mu$ is a float number in the interval $[0,1]$ produced by a standard random number generator:

$$\Delta m = \mu \, (m^t_i + m^t_j \,)/2$$

$$m^{t+1}_i = m^t_i - \Delta m$$

$$m^{t+1}_j = m^t_j + \Delta m \qquad\qquad (1)$$

This particular rule of trade is selected for simplicity and extensive use. Comparisons can be established with popularly referenced literature [4-9]. Here, the transaction of money is quite asymmetric as agent $j$ wins the money that $i$ losses. Also, if agent $i$ has not enough money $(m^t_i < \Delta m)$, no transfer takes place.

The chaotic selection of agents $(i, j)$ for each interaction, is obtained by a particular 2D chaotic system, the Logistic bimap, described by López-Ruiz and Pérez-García in [17]. This system is given by the following equations and depicted in Fig.1:

$$T:[0,1]\times[0,1] \rightarrow [0,1]\times[0,1]$$

$$x_t = \lambda_a \, ( \, 3 \, y_{t-1} + 1) \, x_{t-1} \, ( \, 1 - 3 \, x_{t-1} )$$

$$y_t = \lambda_b \, ( \, 3 \, x_{t-1} + 1 \, ) \, y_{t-1} \, ( \, 1 - 3 \, y_{t-1} ) \qquad\qquad (2).$$

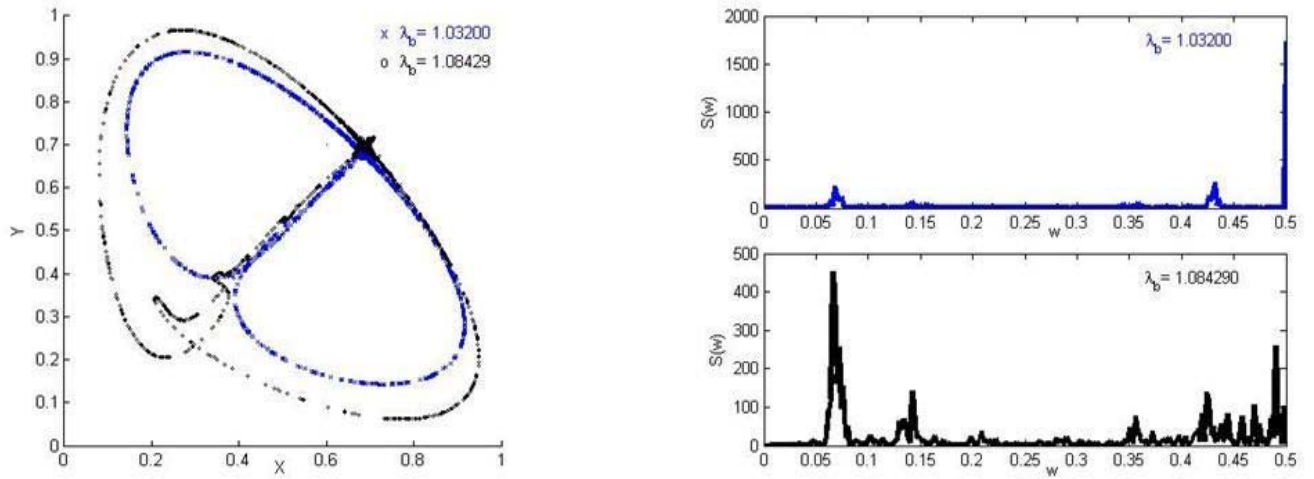

**Fig. 1 (Left) Representation of 2000 points of the chaotic attractor of the Logistic bimap. This bimap is symmetric respective to the diagonal y=x when $\lambda_a = \lambda_b$ . (Right) Representation of the spectrum of coordinate $x_t$ .The values of the parameters for both graphics are $\lambda_a = \lambda_b = 1.032$ in the blue curves and $\lambda_a=1.032$, $\lambda_b=1.08429$ in the black curves. The curves show the symmetric case, $\lambda_a = \lambda_b = 1.032$, and the most asymmetric one while chaotic behaviour is still observed, $\lambda_a=1.032$, $\lambda_b=1.08429$.**

The chaotic system of Eq. (2) considers two Logistic maps that evolve in a coupled way. This system is used to select the interacting agents at a given time. The pair **(i , j)** is easily obtained from the coordinates of a chaotic point at instant **t**, $\mathbf{X_t =[x_t, y_t]}$, by a simple float to integer conversion:

$$i = (int) ( x_t * N )$$

$$j = (int) ( y_t * N ) \tag{3}$$

This procedure will make that the selection of a winner **(j),** or a looser **(i),** follows a chaotic pattern, that of the Logistic map. Moreover, as there are two Logistic maps are coupled with the other, this bimap is able to introduce a strong correlation in the selection of agents of each group. This correlation is modulated or governed by the parameters **λ**. The Logistic bimap presents a chaotic attractor in the interval **$\lambda_{a,b}$ €** **[1.032, 1.0843]** (see a real-time animation in [18]). Consequently, the chaotic selection of agents is guaranteed by taking some appropriate **λ** in this range. Fig.1 (left) shows two trajectories for two different groups of values of **$\lambda_a$ and $\lambda_b$**, showing maximum asymmetry while prevailing chaotic behaviour.

An interesting property of the Logistic bimap is that it is symmetric respective to the diagonal **y = x** , when $\lambda_a = \lambda_b$ (see Fig. 1 (left)). The spectrum of coordinate $x_t$ shows a peak for $\omega = 0.5$ (see Fig. 1 (right)) presenting an oscillation of period two that makes a jump over the diagonal axis alternatively between consecutive instants of time. Both sub-spaces **x>y** and **y>x** are visited with the same frequency and the shape of the attractor is symmetric. When $\lambda_b$ becomes greater than $\lambda_a$ the part of the attractor in sub-space **x<y** becomes wider and the frequency of visits of each sub-space becomes different. These characteristics can be used in the model as an input variable, when the degree of symmetry in the selection of agents is properly modified.

In the end, the dynamics of the economic interactions are going to be complex and quite different from the random scenario, where any agent may interact with any other with the same probability. Here, market transactions are restricted in the sense that one agent will only interact with specific groups of other agents. To see this, just think that an $i_0$ agent given by an **x=x_0** coordinate, with $x_0$ any constant value in the interval **[0,1]**. Draw the line **x=x_0** in Fig.1 (left) and then it will be seen that this agent will interact with its neighbours on the diagonal **(j ≈ i)** and maybe with two other distant groups of agents. Additionally these secondary groups interact with other groups and so on, giving rise to a complex flow of interactions in the whole market. This may seem more realistic than the random case, as normally one individual doesn't interact directly with any other, but with specific groups of traders, being the complex connexions of the community, the ones that develop a global indirect trading.

## 3.  SIMULATION RESULTS OBTAINED IN THE CHAOTIC GAS-LIKE MODEL

Taking into account the chaotic market model discussed in the previous section, different computer simulations are carried out. In these simulations a community of **N=5000** agents with initial money of $\mathbf{m_0}$**=1000\$**  is taken. The simulations take a total time of $\mathbf{T=2*N^2}$ **= 50** millions of transactions.

Different cases are produced when different values of the chaotic parameters $\lambda_a$ and $\lambda_b$. In this way the symmetry of the selection of agents can be varied at will. This will allow us to see the effects of introducing an asymmetry in the process of selection of interacting agents. Simulations are produced when $\lambda_a = 1.032$ has a fixed value and $\lambda_b$ varies from **1.032** to **1.033162** with increments of $\delta = 0.581 \times 10^{-4}$.

The results of these simulations reveal very interesting features. First, there is a group of individuals that keep their initial money and don't interact at all. This result would be consistent with real markets where not every agent are active. This number is due to the shape of the chaotic attractor. In Fig. 1 (Left), it can be seen that the attractor hardly reaches the extreme values of the interval **[0,1]**. In the case of $\lambda_a = \lambda_b = 1.032$ the number of non interacting agents is **1133** in a community of **N=5000**. When $\lambda_b$ increases the attractor expands and this number becomes smaller.

After removing passive agents and their money, the final distributions of money are shown in Fig. 2 with different axes scales and for different values of $\lambda_b$. Here $\lambda_a = 1.032$ for all cases. As it can be seen, the symmetric case produces an exponential distribution, as in the random scenario of the gas-like model [5]. Amazingly, increasing the asymmetry of the chaotic selection, the money distribution degenerates progressively from the exponential shape to a power law profile.

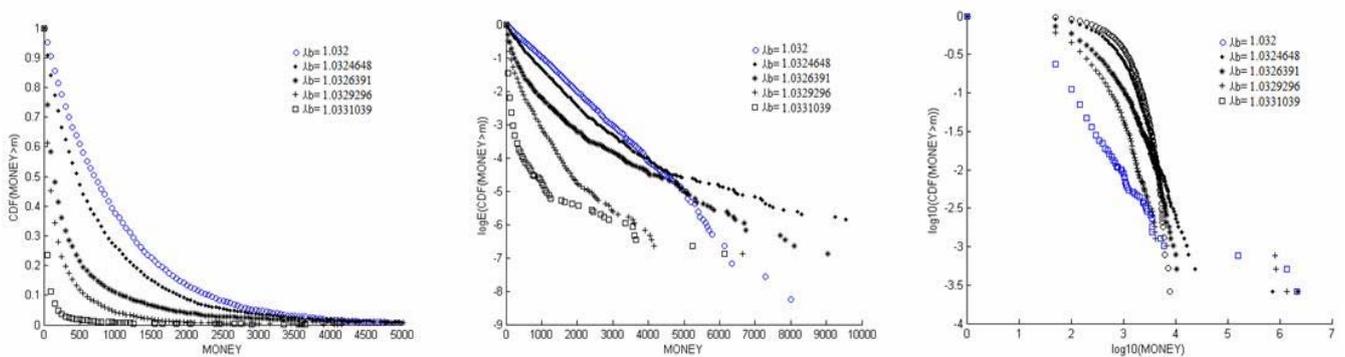

**Fig.2 (Left) Final CDF's obtained for parameters $\lambda_a = 1.032$ and $\lambda_b = 1.032, 1.324648, 1.326391, 1.329296$ and $1.331039$. (Middle) Representation of the same CDF's in a logarithmic plot. (Right) representation of the same CDF's in a double logarithmic plot.**

Fig. 2 (Left) shows in five curves, the cumulative distribution functions (CDF's) obtained for the following four cases: $\lambda_a$ = **1.032** **and** $\lambda_b$ = **1.032, 1.324648, 1.326391, 1.329296** and **1.331039**. Here the probability of having a quantity of money bigger or equal to the variable **MONEY**, is depicted in natural axis plot. The symmetric case $\lambda_a$ = $\lambda_b$ =**1.032** is highlighted in blue colour. Fig. 2 (Middle) shows the same data in a natural logarithmic plot, to illustrate the exponential profile obtained in the symmetric case (highlighted in blue). It also shows how a progressive asymmetry in the selection of agents degrades this characteristic profile. Fig. 2 (Right) shows again the same data but in a double-logarithmic plot. It gives an extensive view of the degradation observed before, making evident the straight line power-law profile obtained in the most asymmetric case (highlighted in blue).

As a consequence, one may say that a chaotic selection of agents is reproducing the two characteristic distributions observed in real wealth distributions. On one hand, one obtains the exponential distribution, as in the random gas-like models. This characteristic appears when chaotic market is operating under a symmetric rule of selection of interacting agents. The significance may be that the correlations in agent interactions looks like random, and all agents win or lose with fair probability.

On the contrary, when asymmetry is introduced in the chaotic process of selection of agents, the distribution of money becomes progressively more unequal. The probability of finding an agent in the state of poorness increases and only a minority of agents reach very high fortunes. What is happening here is, that the asymmetry of the chaotic map is selecting a set of agents preferably as winners for each transaction (**j** agents). While others, with less chaotic luck, become preferably the losers (**i** agents). As in real life, there are markets where some individuals possess a preferred status and this makes them win in the majority of their transactions.

From these primary results, it seems interesting to study the dynamics of the system at micro detail. This might help to uncover the possible causes that originate unfair distributions of money in society. To do that, simulations are repeated for a more tractable number of agents **N=500.** The initial money is **$m_0$=1000\$** as previously, and the simulations take a total time of **T=2\*$N^2$ = 0.5** millions of transactions.

First the CDF is obtained for three simulation cases with different **$\lambda_b$** values (see Fig. 3 (Left)). The degradation of the exponential distribution is again obtained as asymmetry increases.

However when Fig. 2 (Left) and Fig. 3 (Left) are compared an avalanche effect becomes now evident for bigger markets. Here one may appreciate that the shape of the money distributions in the symmetric case is the same for a simulation scenario of **N=5000** or **N=500** agents. In contrast, the asymmetric cases produce more unequal or dramatic distributions when the community of agents is greater. A slight change in the asymmetry of the interactions, compare for example the case **$\lambda_b$=1.033162,** produces a significantly higher degradation in the final distribution of money for the **N=5000** market. At this point, it is interesting to remark that, in a chaotic scenario the size of the market matters. A small asymmetry in the rules of selection of agents will become far more aggressive in its final distribution of wealth depending on the size of the market.

As a consequence, the globalization of markets that evolve under chaotic rules, may suffer avalanche effects that drive more unequal and dramatic distributions of wealth. This may resemble some situations observed in real economy, where small unbalances in global markets, may produce large differences in the share of capital among individuals.

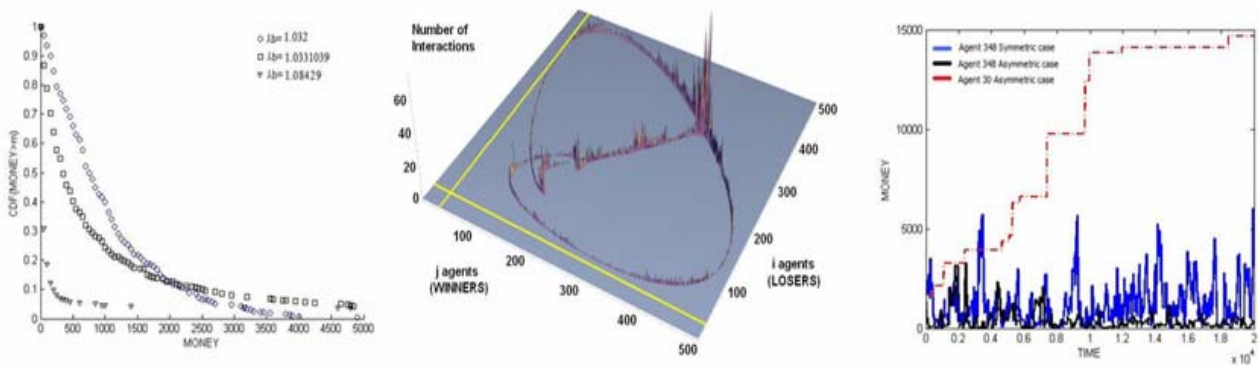

**Fig. 3 Results obtained for a simulation scenario with N=500 agents. (Left) Representation of the same CDF's of Fig. 2 for N=500 agents and three different cases ($\lambda_b$=1.032, 1.033162 and 1.08429). (Middle) Number of transactions between pairs of agents for the most asymmetric case ($\lambda_a$=1.032 and $\lambda_b$=1.08429). Axis x and y represent pairs of agents (j,i), selected as winner (j) or loser (i) respectively. Axis z represents the number of transactions of each pair. Agent 30 is highlighted in yellow to show that he never looses (i agent). (Right) Time evolution of money for agent 348 in the symmetric case ($\lambda_a$=$\lambda_b$=1.032), and agents 348 and 30 in the most asymmetric case ($\lambda_b$=1.08429).**

Secondly, the network of chaotic economic interactions is analyzed. Agents trade in a complex network of interactions illustrated in Fig. 3 (Middle). As we can see, agents in the range **320** to **360** have quite a higher number of interactions between them. Amazingly, these agents are not specially treated in the symmetric case, for in this scenario all agents have the same opportunities. On the other hand, the figure shows how in the most asymmetric case some agents are preferably selected as winners. The maximum asymmetry case depicted here treats some agents as winners in all transactions. See highlighted in yellow, agent **30**, who is never selected as an **i** agent, a looser. In this situation the group of the agents in the range **320** to **360**, which interact mainly among themselves, will always get poorer as the majority of the society is getting poor. The more they interact the sooner they will lose all their money. Put it in another way, in the asymmetric case, not all agents have the same opportunities.

In third place, Fig.3 (Right) shows the evolution in time of an agent's money in symmetric and asymmetric cases. Here agent number **348,** who in the range **320** to **360,** is depicted as an example.

In the picture it can be observed that when the market is symmetric, agents' wealth oscillates in a random-like way. Becoming rich or poor in the end it is just a question of having a very profitable transaction, being in the right place at the right time. In the chaotic market model this is simply the result of some specific selection of the initial conditions and chaotic parameters. Let us note that in the symmetric case, the money distribution becomes exponential. On a global perspective, there are no sources or sinks of wealth and even then, when the individuals have the same opportunities, the final distribution is unequal.

However when the market is asymmetric agent **348** becomes inevitably poorer as time passes. In this case there are agents that never lose (as for example, agent **30)** and they become inevitably sinks of wealth. This can be observed in Fig.3 (Right) where agent **30** is precisely one of these sinks, accumulating more and more money every time he trades. These chaotic favoured agents make the rest of the community poorer. On a global perspective, becoming rich in the asymmetric case depends on who you are and who you interact with (the shape of the chaotic attractor and its initial conditions). This resembles a society where some individuals belong to specific circles of economic power.

Finally, to illustrate these qualitative observations, we study the number of times that agents win or lose in the simulation cases. The results obtained are depicted in Fig. 4. Here it is shown the number of times (interactions) that an agent has been a looser (bottom graph) and the difference of winning over losing times (top graph). The **x** axis shows the ranking of agents arranged in descending order according to their final wealth. So, agent number **0** is the richest of the community and agent number **500** is the poorest.

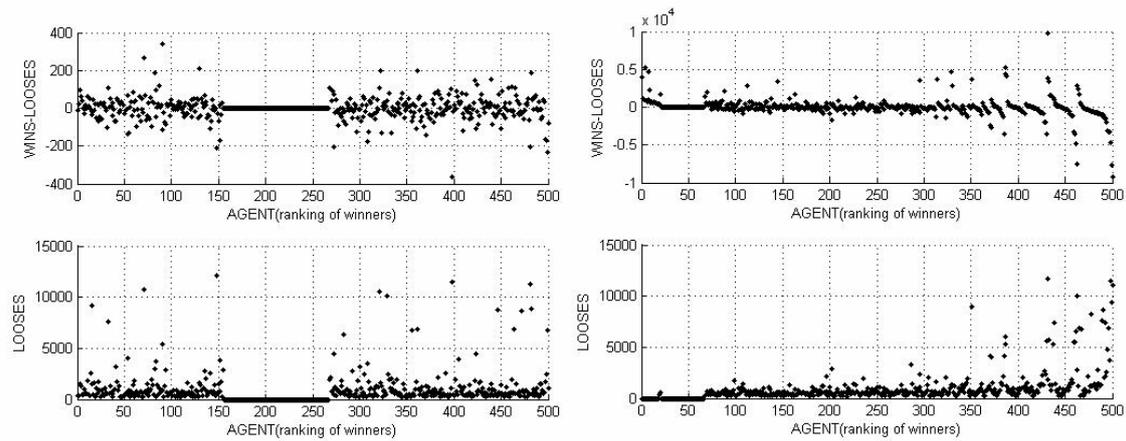

**Fig. 4 Representation of the role of N=500 agents in the community after all interactions. Agents are arranged in descending order according to their final wealth (agent 0 is the richest of all). The upper graphic shows the total number of wins over looses of an agent. The bottom graphic shows the number of times an agent has been selected as i agent or looser. (Left) Symmetric case with $\lambda_a = \lambda_b =1.032$. Here, the number of wins and looses is uniformly distributed among the community. A range of 112 individuals are passive and always have the initial 1000$. (Right) Asymmetric case with $\lambda_a = 1.032$ and $\lambda_b = 1.084290$. Here, 46 agents never interact and 18 agents never loose.**

Fig. 4 (Left) depicts the symmetric case, where $\lambda_a = \lambda_b =1.032$. Here, the number of wins and looses is uniformly distributed among the community. There is also a range of agents that don't interact (**112** agents), this can be seen clearly in the figure now. In this case, the chaotic selection of agents shows no particular preference for any other and the final distribution becomes the exponential. This is similar to traditional gas-like simulations with random agents [4-9].

Fig. 4 (Right) shows the same magnitudes for the asymmetric case with $\lambda_a = \mathbf{1.032}$ and $\lambda_b = \mathbf{1.0842908}$. Here the asymmetry is maximum prevailing chaotic behaviour. In this case, it can be seen that there is a group of agents in the range of maximum richness that never loose. The chaotic selection is giving them maximum luck and this makes them richer and richer at every transaction. A range of agents, lower than in the symmetric case, are passive and never interact (**46** agents).

In this case, the final money distribution becomes quite unequal: **269** agents (half of the society approximately) become in state of poorness with a final wealth inferior to 500$ and of them, and **56** agents (approx. the 10%) finish with less that 50$.

It is also interesting to see in Fig. 4 (Right) that in the poor class, there are agents that have a positive difference of wins over looses, but amazingly they become poor anyway. Consequently, one can deduce that they are also bad luck individuals. They may be selected as **j** agents in most part of their transactions, but unfortunately their corresponding trading partners (**i** agents) are poor too, and they can effectively earn low or no money in these interactions.

To summarize, these results show that the symmetric case of selection of agents resembles a society where agents have equal opportunities in the market. In this case, agents are equally selected winners or losers (symmetry, **x↔y** and **i↔j**) in a chaotic and correlated fashion. This resembles a market where all agents have the same opportunities of success. In this case, there is no particular preference for any other and the final distribution of money becomes exponential, as in the random gas-like model.

Quite the opposite, when a small asymmetry is introduced in the rule of selecting agents, the society becomes more unequal. Some agents become preferred winners for most of their transactions and the money. The opportunities of winning in the end just depend on interacting with the proper group of rich agents. This resembles a market where there are established circles of power and corruption.

Additionally, the chaotic nature of the market may also reproduce an avalanche effect as a result of globalization of markets. In this case, a small asymmetry may produce extreme inequality in the final distribution of money.

## 4. CONCLUSIONS

This work introduces a novel approach in the field of economic (ideal) gas-like models [4-9]. It proposes a chaotic selection of agents in a trading community and it accomplishes to obtain the two most important money distributions observed in real economies, exponential and power-law distributions, besides a mechanism of transition between them.

The introduction of chaos is based on two considerations. One is that real economy shows a kind of a chaotic character. The other one is that dynamical systems offer a very simple and flexible tool able to break the inherent symmetry of random processes and overcome the technical difficulties of gas models to shift from the Boltzmann-Gibbs to the Pareto distributions.

A specific 2D dynamical system under chaotic regime is considered to produce the chaotic selection of agents. This system represents two logistic maps coupled in a multiplicative way and produces a strong correlation between the interacting agents. This correlation guaranties a complex behaviour that can be tuned through two system parameter ( $\lambda_a$ and $\lambda_b$ ) and so, it is able to produce symmetric or asymmetric conditions in the market.

The results obtained in this model show how a chaotic selection of agents under symmetric conditions produces a final exponential distribution of money. In contrast, when asymmetry is introduced in the selection of agents, the distribution of money becomes progressively more unequal until it produces power law profiles.

The analysis of micro level interactions in both cases shows what is happening in the market and what is responsible of producing these two global phases (exponential or power law).

The main conclusions produced by this model may resemble some situations in real economy. In a symmetric scenario the selection of agents resembles a society where agents have equal opportunities in the market. In this case, there is no particular preference for any other and the final distribution becomes the exponential. This is similar to traditional gas-like simulations with random agents. On a global perspective, there are no sources or sinks of wealth and even when the individuals have the same opportunities, the final distribution is unequal. From an agent point of view, becoming rich or poor just depends of being at the right time in the right place.

In the asymmetric scenarios, a small asymmetry is introduced in the rule of selecting agents and then, the society becomes unequal. Some agents become preferred winners for most of their transactions. The opportunities of winning in the end just depend on being a preferred agent and interacting with the proper group of rich agents. The final money distribution resembles the Pareto distribution. Moreover, the chaotic nature of the market also reproduces the impact of the size of the market on the final distribution of wealth. When unequal or asymmetric conditions of trading are ruling, an avalanche effect may be produced and then inequality becomes extreme.

The authors hope that this work that may bring new ideas and perspectives to Economy and Econophysics. The proposal of considering chaotic dynamics in multi-agents modelling may also be of interest to other fields, where scientists try to describe and understand complex systems.

Acknowledgements The authors acknowledge some financial support by Spanish grant DGICYT-FIS2009-13364-C02-01.